\documentclass[%
 aip,
 amsmath,amssymb,
 reprint,%
]{revtex4-2}

\usepackage{graphicx}
\usepackage{dcolumn}
\usepackage{bm}

\usepackage{float}
\usepackage[utf8]{inputenc}
\usepackage[T1]{fontenc}
\usepackage{mathptmx}
\usepackage{etoolbox}
\usepackage{bbold}
\usepackage{color}
\usepackage{pgfplots}
\pgfplotsset{compat=newest}
\usepackage{tikz}
\usetikzlibrary{patterns}

\makeatletter
\def\@email#1#2{%
 \endgroup
 \patchcmd{\titleblock@produce}
  {\frontmatter@RRAPformat}
  {\frontmatter@RRAPformat{\produce@RRAP{*#1\href{mailto:#2}{#2}}}\frontmatter@RRAPformat}
  {}{}
}%
\makeatother

\makeatletter
\newcommand{\thickhline}{%
    \noalign {\ifnum 0=`}\fi \hrule height 1pt
    \futurelet \reserved@a \@xhline
}
\newcolumntype{"}{@{\hskip\tabcolsep\vrule width 1pt\hskip\tabcolsep}}
\makeatother

\begin{document}

\preprint{AIP/123-QED}

\title[Driven damped harmonic oscillator revisited: energy resonance]{Driven damped harmonic oscillator revisited: energy resonance}

\author{K. Lelas}
\email{klelas@ttf.unizg.hr}
\affiliation{Faculty of Textile Technology, University of Zagreb, Croatia}
\author{N. Poljak}
\email{npoljak.phy@pmf.hr}
\affiliation{Department of Physics, Faculty of Science, University of Zagreb, Croatia}

\date{\today}

\begin{abstract}
We derive the exact expression for the resonant frequency of the time-averaged steady-state energy and show that this frequency is excellently approximated by the arithmetic mean of the amplitude and velocity resonant frequencies. In addition, we argue that the frequency of the amplitude resonance can be regarded as the energy resonance frequency, since it provides the maximal peak values of instantaneous energy. 
\end{abstract}

\maketitle

\section{Introduction}

The driven damped harmonic oscillator is regularly covered in standard undergraduate physics textbooks \cite{UniPhys, Resnick10, Berkeley, Morin1, Morin2, Dourmashkin}, and in the context of resonance phenomena, amplitude resonance and/or velocity (power) resonance are typically considered. In some textbooks \cite{ Berkeley, Morin1, Morin2, Dourmashkin}, energy and the time-averaged energy in steady-state are analyzed, but energy resonance is not discussed at all, while time-averaged energy resonance is considered rarely and only in the weak damping limit. 
In this note, we address the resonant behavior of energy in detail. In addition, we comment on some issues with the presentation of the input power and power loss in standard textbooks, as well as on the steady-state acceleration resonance. 


As an example of a driven damped harmonic oscillator, we consider a block of mass $m$ that oscillates under the influence of three forces, i.e.\ the restoring force of a spring $F_{res}(t)=-kx(t)$, where $k$ is the stiffness of the spring, the damping force $F_d(t)=-b\dot x(t)$, where $b>0$ is the damping constant, and the external driving force $F_{ext}(t)=F_0\cos(\omega t)$, where $\omega$ is the angular frequency of the driving force. The associated equation of motion, i.e.\ $m\,\ddot x(t)=F_{res}(t)+F_d(t)+F_{ext}(t)$, can be rewritten in the form    
\begin{equation}
\ddot x(t)+2\gamma\dot x(t)+\omega_0^2x(t)=\frac{F_0}{m}\cos(\omega t)\,,
\label{DHOeq}
\end{equation}
where $\gamma=b/(2m)>0$ is the damping coefficient and $\omega_0=\sqrt{k/m}$ is the natural angular frequency of the system.

Standard physics textbooks for undergraduate studies \cite{Berkeley,Morin1,Morin2,Dourmashkin} thoroughly discuss how the steady-state behavior of the damped driven harmonic oscillator is determined by the particular solution of nonhomogeneous differential equation \eqref{DHOeq}, i.e.\ the steady-state solution of equation \eqref{DHOeq} is given by 
\begin{equation}
x_s(t)=A(\omega)\cos(\omega t-\varphi)\,,
\label{steady}
\end{equation}
where 
\begin{equation}
A(\omega)=\frac{F_0}{m\sqrt{(\omega_0^2 - \omega^2)^2+(2\gamma \omega)^2}}
\label{amplitude}
\end{equation}
is the amplitude, and the phase-lag $\varphi$ is given by
\begin{equation}
\tan \varphi = \frac{2\gamma\omega}{\omega_0^2 - \omega^2}\,.
\label{phase}
\end{equation}
%

\section{Amplitude resonance and velocity resonance}

In this section, we give a short review of the amplitude resonance and the velocity resonance, i.e.\ a short review of well-studied resonances, and we comment on some minor issues with the treatment of power resonance in the literature. The driving frequency that maximizes the steady-state displacement amplitude \eqref{amplitude} is \cite{Morin1} 
%
%
%
\begin{equation}
\omega_A=\sqrt{\omega_0^2-2\gamma^2}\,.
\label{omegaA}
\end{equation}
%
We can see from \eqref{omegaA} that amplitude resonance can only occur if $\omega_0^2-2\gamma^2>0$ is true. Hence, this type of resonance is limited to underdamped systems with $\gamma<\sqrt{2}\omega_0/2$, while amplitude resonance effects are never present in systems with $\gamma\geq\sqrt{2}\omega_0/2$. Taking into consideration \eqref{phase} and \eqref{omegaA} we can easily deduce
that the phase-lag angle $\varphi (\omega_A)$ depends on the amount of damping in the system and takes on a range of values, i.e.\ $\varphi (\omega_A)\in(0,\pi/2)$, where the limits of the range are $\varphi(\omega_A)\rightarrow\pi/2$ as $\gamma\rightarrow0$ and $\varphi(\omega_A)\rightarrow 0$ as $\gamma\rightarrow\sqrt{2}\omega_0/2$. 
%
%
%
%


The velocity corresponding to the steady-state solution \eqref{steady} is given by
\begin{equation}
v_s(t)=\dot{x}_s(t)=-V(\omega)\sin(\omega t-\varphi)\,,
\label{Vsteady}
\end{equation}
where $V(\omega)=\omega A(\omega)$ is the velocity amplitude. 
%
%
The driving frequency that maximizes the velocity amplitude is \cite{Resnick10} 
%
%
%
\begin{equation}
\omega_V=\omega_0 \,.
\label{omegaV}
\end{equation}
%
We see that $\omega_V$ does not depend on damping, and thus, contrary to the amplitude resonance, velocity resonance effects are present for any value of the damping coefficient $\gamma$. We should note that in this type of resonance the phase-lag angle is $\varphi(\omega_V)=\pi/2$ for any damping coefficient $\gamma$, since $\tan \varphi(\omega_V) \rightarrow \infty$ regardless of the value of $\gamma$. Thus, only in the weak damping limit are both resonant frequencies and phase-lag angles approximately equal, i.e. $\omega_A\approx\omega_V$ and $\varphi(\omega_A)\approx\varphi(\omega_V)$ only if $\gamma\ll\omega_0$. 
%
%
%
%

The input power and the power loss are important quantities to consider in the analysis of a driven damped oscillator\cite{Berkeley}. The steady-state instantaneous input power is given by
\begin{align}
P_{in}(\omega,t)&=F_{ext}(t)\, v_s(t) \nonumber \\
&=-F_0\omega A(\omega)\cos(\omega t)\sin(\omega t-\varphi)\,,
\label{Pin}
\end{align}
while the steady-state instantaneous power loss is given by 
\begin{align}
P_{out}(\omega,t)&=F_d(t)\, v_s(t) \nonumber \\
&=-2m\gamma\,\omega^2 A^2(\omega)\sin^2(\omega t-\varphi)\,.
\label{Ploss}
\end{align}
Denoting the time average over one period by brackets, i.e.\ $\langle f(t)\rangle=T^{-1}\int_t^{t+T}f(t')\,\textrm{d}t'$, where $T=2\pi/\omega$ is the period of the driving force, we get $\langle\cos(\omega t)\sin(\omega t-\varphi)\rangle=-\sin(\varphi)/2$  
%
%
and
%
%
$\langle\sin^2(\omega t-\varphi)\rangle=1/2$. Thus, 
\begin{equation}
\langle P_{in}(\omega)\rangle=\frac{F_0\omega A(\omega)\sin(\varphi)}{2}=m\gamma\,\omega^2 A^2(\omega)\,
\label{Pinav}
\end{equation}
is the time-averaged steady-state input power, and 
\begin{equation}
\langle P_{out}(\omega)\rangle=-m\gamma\,\omega^2 A^2(\omega)\,
\label{Plossav}
\end{equation}
is the time-averaged steady-state power loss. We used the known identity $\sin(\arctan(\alpha))=\alpha/\sqrt{1+\alpha^2}$, i.e. $$\sin(\varphi)=\frac{2\gamma\omega}{\sqrt{(\omega_0^2-\omega^2)^2+(2\gamma\omega)^2}}\,,$$ when simplifying \eqref{Pinav}. The time-averaged input power \eqref{Pinav} is equal in magnitude to the time-averaged power loss \eqref{Plossav} for any driving frequency $\omega$ and damping coefficient $\gamma$, i.e. $\langle P_{out}(\omega)\rangle=-\langle P_{in}(\omega)\rangle$ for all $\omega>0$ and $\gamma>0$. We note here that the expression for $\langle P_{out}(\omega)\rangle$ in some physics textbooks \cite{Berkeley} is missing the minus sign and, in addition, it is stated several times that $\langle P_{in}(\omega)\rangle=\langle P_{out}(\omega)\rangle$, which could potentially be confusing to students. It is easy to conclude that $\omega_V$ maximizes the amplitude of instantaneous power loss, i.e.\ the peak value of \eqref{Ploss}, as well as the magnitudes of the time-averaged input power \eqref{Pinav} and power loss \eqref{Plossav}, since all three quantities are proportional to $V^2(\omega)=\omega^2A^2(\omega)$. What is not obvious at first glance, and is not stated even in advanced undergraduate level textbooks \cite{Berkeley, Morin1, Morin2, Dourmashkin}, is that the frequency $\omega_V$ also maximizes the peak value of instantaneous input power \eqref{Pin}. 
We argue in a simple way why this is indeed the case. The function \eqref{Pin} has two factors that depend on $\omega$, i.e. $\omega A(\omega)$ and the product $-\cos(\omega t)\sin(\omega t-\varphi)$, where $\varphi$ also depends on $\omega$. The first factor alone has maximum for $\omega=\omega_V$, while the second factor has maximal peak value if $\varphi=\pi/2$, i.e.\ also for $\omega=\omega_V$, since in that case $-\cos(\omega t)\sin(\omega t-\pi/2)=\cos^2(\omega t)$, so we can conclude that $\omega_V$ maximizes the peak value of \eqref{Pin}. Thus, we can refer to $\omega_V=\omega_0$ as the resonant frequency of input power and power loss, or simply as the power resonance frequency $\omega_P=\omega_0$.       

\section{Energy resonance}

The total oscillator steady-state instantaneous energy is
\begin{equation}
E(\omega,t)=\mathcal{E}_P(\omega)\cos^2(\omega t-\phi)+\mathcal{E}_K(\omega)\sin^2(\omega t-\phi)\,,
\label{totalE}
\end{equation}
where
\begin{equation}
\mathcal{E}_P(\omega)=\frac{m\omega_0^2A^2(\omega)}{2}
\label{ampEP}
\end{equation}
is the amplitude of the potential energy, and 
\begin{equation}
\mathcal{E}_K(\omega)=\frac{m\omega^2A^2(\omega)}{2}
\label{ampEK}
\end{equation}
is the amplitude of the kinetic energy. 
Since $\mathcal{E}_P(\omega)/\mathcal{E}_K(\omega)=\omega_0^2/\omega^2$, for driving frequencies $\omega<\omega_0$ we have $\mathcal{E}_P(\omega)>\mathcal{E}_K(\omega)$, i.e.\ the total energy \eqref{totalE} is not constant, but rather oscillates with peak values equal to $\mathcal{E}_P(\omega)$. For $\omega=\omega_0$, i.e.\ at the velocity resonance frequency, $\mathcal{E}_P(\omega_0)=\mathcal{E}_K(\omega_0)$ and the total energy is constant. Since $\textrm{d}E(\omega,t)/\textrm{d}t=P_{in}(\omega,t)+P_{out}(\omega,t)$, and $\textrm{d}E(\omega,t)/\textrm{d}t=0$ for $\omega=\omega_0$, the instantaneous input power and power loss are in balance, i.e. $P_{out}(\omega_0,t)=-P_{in}(\omega_0,t)$ at the velocity resonance. Thus, for $\omega=\omega_0$ the system behaves as an undamped free oscillator that oscillates with total energy $E(\omega_0,t)=m\omega_0^2A^2(\omega_0)/2$. We note here that it is stated in textbook \cite{Berkeley} that the input power and power loss are equal in magnitude only if averaged over one cycle, which is not true, as we just commented. For driving frequencies $\omega>\omega_0$ we have $\mathcal{E}_P(\omega)<\mathcal{E}_K(\omega)$, i.e.\ the total energy \eqref{totalE} oscillates with peak values equal to $\mathcal{E}_K(\omega)$. Note that this behavior of energy \eqref{totalE} is valid for any value of the damping coefficient $\gamma$. The resonant frequencies $\omega_A$ and $\omega_V$ are also resonant frequencies of the amplitudes of potential and kinetic energy, i.e.\ of \eqref{ampEP} and \eqref{ampEK}, respectively. The resonant amplitude of potential energy as a function of $\gamma$ is
\begin{equation}
\mathcal{E}_P(\omega_A)=\frac{F_0^2}{8m\gamma^2\left(1-(\gamma/\omega_0)^2\right)}\,,
\label{Ep1}
\end{equation}
and the resonant amplitude of kinetic energy as a function of $\gamma$ is
\begin{equation}
\mathcal{E}_K(\omega_V)=\frac{F_0^2}{8m\gamma^2}\,.
\label{Ek1}
\end{equation}
The relation \eqref{Ep1} is valid for $\gamma\leq\sqrt{2}\omega_0/2$. We see that $\mathcal{E}_P(\omega_A)>\mathcal{E}_K(\omega_V)$ is valid, and due to the discussion in the previous paragraph we can conclude that the peak values of energy \eqref{totalE} are maximized for $\omega=\omega_A$. In that sense we can say that the amplitude resonance frequency is also the \emph{energy resonance frequency} $\omega_E$, i.e.
\begin{equation}
\omega_E=\omega_A=\sqrt{\omega_0^2-2\gamma^2}\,.
\label{omegaE}
\end{equation}
To the best of our knowledge, this view on the resonance of total instantaneous energy is not taken anywhere else in the literature. 

\begin{figure}[h!t!]
\begin{center}
\includegraphics[width=0.48\textwidth]{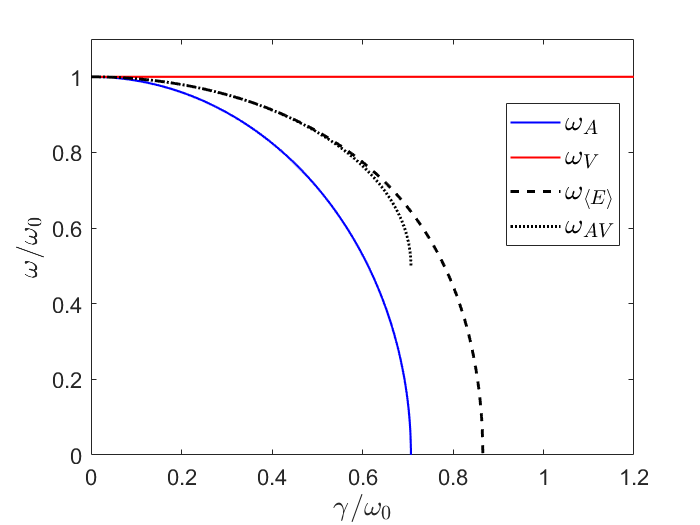}
\end{center}
\caption{Resonant frequencies $\omega_A$ (solid blue curve), $\omega_V$ (solid red line), and $\omega_{\langle E\rangle}$ (dashed black curve) as functions of the damping coefficient $\gamma$. In addition, we show $\omega_{AV}=(\omega_A+\omega_V)/2$ (dotted black curve) and we can see that $\omega_{AV}$ excellently approximates $\omega_{\langle E\rangle}$ for a wide range of damping coefficients, i.e. for $\gamma\lesssim0.65\omega_0$. The solid blue and dotted black curves end at the points $(\gamma,\omega_A)=(\sqrt{2}\omega_0/2,0)$ and $(\gamma,\omega_{AV})=(\sqrt{2}\omega_0/2,\omega_0/2)$ respectively, while the dashed black curve ends at point $(\gamma,\omega_{\langle E\rangle})=(\sqrt{3}\omega_0/2,0)$. The approximation $\omega_{AV}$ is not applicable if $\gamma\geq\sqrt{2}\omega_0/2$, since the amplitude resonance does not occur in that case.} 
\label{omegaEE}
\end{figure}

Another important quantity in the analysis of a driven damped oscillator is the energy \eqref{totalE} averaged over one period of the driving force \cite{Berkeley}, i.e.\ the time-averaged steady-state energy given by
\begin{equation}
\langle E(\omega)\rangle=\frac{1}{T}\int_t^{t+T}E(t')dt'=\frac{\mathcal{E}_P(\omega)+\mathcal{E}_K(\omega)}{2}\,.
\label{avE}
\end{equation}
We will refer to \eqref{avE} simply as the \emph{average energy}. We note here that the term $\mathcal{E}_P(\omega)/2$ corresponds to the average potential energy and $\mathcal{E}_K(\omega)/2$ to the average kinetic energy. 
Resonance of the average energy is rarely discussed in the literature, and even then only in the weak damping limit, i.e.\ for $\gamma\ll\omega_0$.\cite{Berkeley, Dourmashkin} Thus, to the best of our knowledge, an exact expression for the resonant driving frequency of the average energy has not been considered in the literature so far. The driving frequency that maximizes \eqref{avE} can be determined from the condition $\frac{\textrm{d}\langle E(\omega)\rangle}{\textrm{d} \omega}\propto\frac{\textrm{d}(\omega_0^2+\omega^2)A^2(\omega)}{\textrm{d} \omega}=0$, and we get
\begin{equation}
\omega_{\langle E\rangle}=\sqrt{2\omega_0\sqrt{\omega_0^2-\gamma^2}-\omega_0^2}
\label{omegaEav}
\end{equation}
for the resonant frequency of the average energy. The value of the average energy at resonance is given by 
\begin{equation}
\langle E(\omega_{\langle E\rangle})\rangle=\frac{F_0^2}{16m\omega_0^2}\cdot\frac{1}{(\gamma/\omega_0)^2-1+\sqrt{1-(\gamma/\omega_0)^2}}\,.
\label{peakEav}
\end{equation}
Using the condition $2\omega_0\sqrt{\omega_0^2-\gamma^2}-\omega_0^2>0$, we determine that the range of possible values of the damping coefficient $\gamma$ for which this resonance occurs is $\gamma\in(0,\sqrt{3}\omega_0/2)$. It is easy to show that $\omega_{\langle E\rangle}\approx\omega_A\approx\omega_V=\omega_0$ in the weak damping limit, i.e.\ for $\gamma\ll\omega_0$, but, in general, we expect $\omega_{\langle E\rangle}$ to occur between $\omega_A$ and $\omega_V$, i.e.\ $\omega_A<\omega_{\langle E\rangle}<\omega_V$, for any $\gamma\in(0,\sqrt{3}\omega_0/2)$, due to the fact that it maximizes the sum of the potential energy amplitude (which is maximal at $\omega_A$) and kinetic energy amplitude (which is maximal at $\omega_V$). In Fig.\,\ref{omegaEE} we clearly see that this is indeed the case.  

We can check analytically how well $\omega_{\langle E\rangle}$ is approximated by $\omega_{AV}=(\omega_A+\omega_V)/2$. The relative difference of $\omega_{AV}$ and $\omega_{\langle E\rangle}$ is
\begin{equation}
R=\frac{\omega_{\langle E\rangle}-\omega_{AV}}{\omega_{\langle E\rangle}} = \left(1 - \frac{1+\sqrt{1-2(\gamma/\omega_0)^2}}{2\sqrt{2\sqrt{1-(\gamma/\omega_0)^2}-1}}\right) \,.
\end{equation}
The Maclaurin series of this expression is missing the first five powers in $\gamma/\omega_0$ and is given by
\begin{equation}
R\approx \frac{1}{16}\left(\frac{\gamma}{\omega_0}\right)^6 + \frac{19}{128}\left(\frac{\gamma}{\omega_0}\right)^8 + ...\,,
\end{equation}
which shows that $R\lesssim1\%$ for $\gamma\lesssim0.65\omega_0$. 

\begin{figure}[h!t!]
\begin{center}
\includegraphics[width=0.48\textwidth]{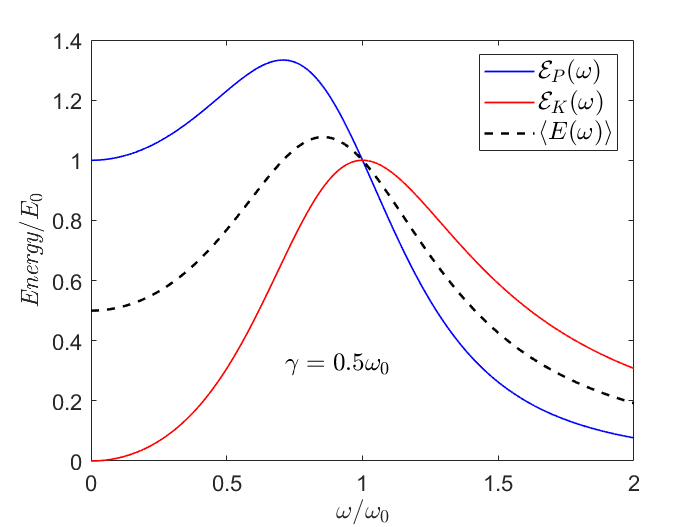}
\end{center}
\caption{The average energy (dashed black curve), the amplitude of the potential energy (solid blue curve) and the amplitude of the kinetic energy (solid red curve), as functions of $\omega$ for $\gamma=0.5\omega_0$. The units on the energy axis are $E_0=F_0^2/(2m\omega_0^2)$.} 
\label{ResonanceAB}
\end{figure}

\begin{figure}[h!t!]
\begin{center}
\includegraphics[width=0.48\textwidth]{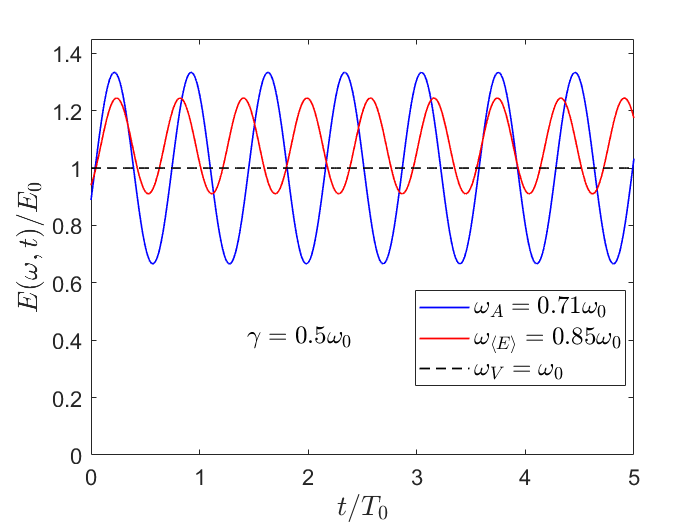}
\end{center}
\caption{The steady-state instantaneous energy \eqref{totalE} for $\gamma=0.5\omega_0$ and driving frequencies $\omega_A$ (solid blue curve), $\omega_{\langle E\rangle}$ (solid red curve) and $\omega_V$ (dashed black horizontal line). The units of energy are $E_0=F_0^2/(2m\omega_0^2)$ and the units of time are $T_0=2\pi/\omega_0$.} 
\label{Etrenutna}
\end{figure}

It is easy to show that the maximal values of the average energy and the amplitudes of potential and kinetic energy are approximately the same, i.e. $\langle E(\omega_{\langle E\rangle})\rangle\approx\mathcal{E}_P(\omega_A)\approx\mathcal{E}_K(\omega_V)=F_0^2/(8m\gamma^2)$, when $\gamma\ll\omega_0$. The differences between individual resonances become significant when $\gamma\ll\omega_0$ is no longer valid. As an example, in Fig.\,\ref{ResonanceAB} we show the average energy and the amplitudes of potential and kinetic energy as functions of $\omega$ for $\gamma=0.5\omega_0$.
In Fig.\,\ref{Etrenutna} we show the steady-state instantaneous energies \eqref{totalE} for driving frequencies $\omega=\lbrace \omega_A, \omega_{\langle E\rangle}, \omega_V\rbrace$ and $\gamma=0.5\omega_0$.    
%
%
%
As we already mentioned, the resonance of average energy was considered in literature only for $\gamma\ll\omega_0$ and $\omega_{\langle E\rangle}\approx\omega_0$ was obtained \cite{Berkeley,Dourmashkin}. We now comment further on quantitative differences that are obtained if we drive the system with $\omega_{\langle E\rangle}$ instead of $\omega_0$. We define ratios $$R_1=\frac{\langle P_{in}(\omega_{\langle E\rangle})\rangle-\langle P_{in}(\omega_0)\rangle}{\langle P_{in}(\omega_0)\rangle}\cdot100\%$$ $$R_2=\frac{\langle E(\omega_{\langle E\rangle})\rangle-\langle E(\omega_0)\rangle}{\langle E(\omega_0)\rangle}\cdot100\%$$ $$R_3=\frac{\langle P_{in}(\omega_{\langle E\rangle})\rangle T(\omega_{\langle E\rangle})-\langle P_{in}(\omega_0)\rangle T(\omega_0)}{\langle P_{in}(\omega_0)\rangle T(\omega_0)}\cdot100\%\,,$$ where $T(\omega)=2\pi/\omega$. For our chosen example, i.e. for a system with $\gamma=0.5\omega_0$, we get $R_1=-8.9\%$, $R_2=7.3\%$, and $R_3=6.4\%$. Thus, if we drive the system with $\omega_{\langle E\rangle}$ instead of $\omega_0$, we get $7.3\%$ higher average energy of steady-state oscillations, with $8.9\%$ smaller time-averaged input power. In terms of input energy per cycle, driving the system with $\omega_{\langle E\rangle}$ requires $6.4\%$ more energy per cycle to maintain steady-state oscillations compared to driving the system with $\omega_0$. 

%
%

\begin{figure}[h!t!]
\begin{center}
\includegraphics[width=0.45\textwidth]{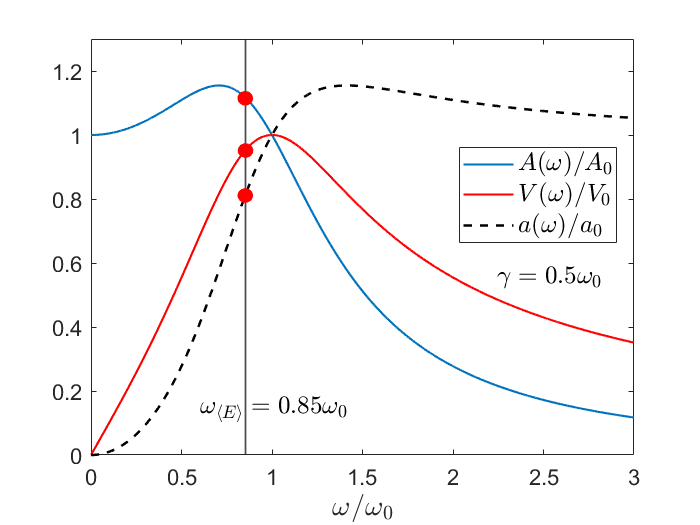}
\end{center}
\caption{Functions $A(\omega)$ (solid bule curve), $V(\omega)$ (solid red curve) and $a(\omega)$ (dashed black curve) for $\gamma=0.5\omega_0$. The filled red circles designate the values of $A(\omega_{\langle E\rangle})$, $V(\omega_{\langle E\rangle})$ and $a(\omega_{\langle E\rangle})$, which correspond to the resonant average energy \eqref{peakEav}. The units used are $A_0=F_0/(m\omega_0^2)$, $V_0=\omega_0A_0$ and $a_0=\omega_0^2A_0$.} 
\label{akceleracija}
\end{figure}

Finally, we note that acceleration resonance is also not considered in standard undergraduate physics textbooks \cite{UniPhys, Resnick10, Berkeley, Morin1, Morin2, Dourmashkin}. The acceleration corresponding to the steady-state solution \eqref{steady} is
\begin{equation}
a_s(t)=\ddot{x}_s(t)=-a(\omega)\cos(\omega t-\varphi)\,,
\label{asteady}
\end{equation}
where $a(\omega)=\omega^2 A(\omega)$ is the acceleration amplitude. The driving frequency that maximizes the acceleration amplitude is easily determined from the condition $\frac{\textrm{d} a(\omega)}{\textrm{d} \omega}=0$
%
which gives
\begin{equation}
\omega_a=\frac{\omega_0^2}{\sqrt{\omega_0^2-2\gamma^2}}=\frac{\omega_V}{\omega_A}\omega_0\,.
\label{omega_a}
\end{equation}
Thus, this resonance occurs at $\omega_a>\omega_0$, for any $\gamma<\sqrt{2}\omega_0/2$. In Fig.\,\ref{akceleracija} we show the scaled displacement amplitude, the scaled velocity amplitude and the scaled acceleration amplitude as functions of $\omega$ for $\gamma=0.5\omega_0$.  

\section{Final remarks}

We would like to point out that, in addition to providing a more detailed theoretical insight into the energy resonances in comparison to standard textbooks, our results can also enrich laboratory exercises in which the driven damped oscillator is considered. For example, in an experimental setup using optical levitation \cite{Levitation}, the parameters corresponding to $F_0$, $m$ and $\gamma$ can be manipulated, values $\gamma\lesssim \sqrt{3}\omega_0/2$ (and higher) can be achieved, and the amplitude $A(\omega)$ can be measured for a range of driving frequencies $\omega$. Using the experimentally obtained $(\omega,A(\omega))$ data, students could calculate the $(\omega,\omega_0^2 A^2(\omega))$, $(\omega,\omega^2A^2(\omega))$ and $(\omega,\omega^2A(\omega))$ data and plot the curves shown in Fig.\,\ref{ResonanceAB} and Fig.\,\ref{akceleracija}. Using the experimentally obtained plots for several different values of $\gamma$, students could gain insight into the behavior of different resonances of the driven damped harmonic oscillator and analyze the validity of the theoretical results presented here. For example, one interesting task could be to check the validity of the approximation $\omega_{\langle E\rangle}\approx\omega_{AV}$ and to determine the values of $A(\omega)$, $V(\omega)$, and $a(\omega)$ that correspond to maximal steady-state average energy for some given $\gamma$. 

We should mention that lab setups which students typically see in their classes when dealing with forced damped oscillations often easily achieve the damping coefficients $\gamma$ which are outside the weak damping limit. Examples to these include, but are not limited, to RLC circuits, systems involving eddy currents and mechanical systems like Pohl's pendulum. Hence, various resonance conditions are easily achievable and students can easily learn about them from experiments.

\section{Acknowledgments}

This work was supported by the QuantiXLie Center of Excellence, a project co-financed by the Croatian Government and European Union through the European Regional Development Fund, the Competitiveness and Cohesion Operational Programme (Grant No. KK.01.1.1.01.0004).\\

\section{Author contributions} 

K.L. and N.P. contributed equally to the paper.

\bibliography{aipsamp}

\end{document}